\documentclass[12pt,english]{article}
\date{}
\usepackage{float}
\usepackage{xcolor}
\usepackage[utf8x]{inputenc}
\usepackage{amsmath}
\usepackage{amsfonts,amsthm}
\usepackage{amssymb}
\usepackage{graphicx,color}
\usepackage{babel}
\usepackage{fontenc}
\usepackage{hyperref}
\usepackage{appendix}
\def\be{\begin{equation}}
\def\ee{\end{equation}}

\begin{document}

\title{{\bf{Holographic information theoretic quantities for Lifshitz black hole}}}

\author{
	{\bf {\normalsize Sourav Karar}$^{a,b}$\thanks{sourav.karar91@gmail.com}},
{\bf {\normalsize Sunandan Gangopadhyay}$^{b}$\thanks{sunandan.gangopadhyay@gmail.com, sunandan.gangopadhyay@bose.res.in }}\\
$^{a}$ {\normalsize Department of Physics, Government General Degree College, Muragachha 741154, Nadia, India}\\
$^{b}${\normalsize Department of Theoretical Sciences, S.N. Bose National Centre for Basic Sciences}\\
{\normalsize Block-JD, Sector III, Salt Lake,  Kolkata 700106, India}
\\[0.3cm]}

\maketitle
\begin{abstract}
\noindent In this paper, we have investigated the holographic entanglement entropy for a linear subsystem in a $3+1$-dimensional
Lifshitz black hole. The entanglement entropy has been analysed in both the infra-red and ultra-violet limits, and has also been computed in the near horizon approximation.
The notion of a generalized temperature in terms of the renormalized entanglement entropy has been introduced. This also leads to a generalized thermodynamics like
law $E=T_g S_{REE}$. The generalized temperature has been defined in such a way that it reduces to the Hawking temperature in the infra-red limit.
We have then computed the holographic subregion complexity. Then the Fisher information metric and the fidelity susceptibility for the same linear subsystem have also been computed using the bulk dual prescriptions. It has been observed that the two metrics are not related to each other. 
\end{abstract}

\newpage

\section{Introduction}

Gauge-gravity duality (\cite{Witten:1998qj}-\cite{Aharony:1999ti}) has been one of the major areas of research in theoretical physics in the 
last two decades. The reason for this intense focus is for its success in explaining the physics
of strongly coupled field theories \cite{Calabrese:2004eu}-\cite{CasalderreySolana:2011us}. The duality makes a connection between a strongly coupled gauge
field theory in $d$- dimensional spacetime and a classical gravity theory in $(d+1)$- dimensional
spacetime, with the field theory living on the boundary of the $(d+1)$- dimensional spacetime.
The advantage of the connection is evident at once. Perturbative calculations which 
could not be performed on the field theory side due to the strong coupling, can now be carried out 
in the gravitational side since the bulk theory is a weakly coupled theory in a classical
gravity background. The duality then translates the information obtained on the gravity side to
a lot of valuable information about the strongly coupled field theories.

The gravitational dual gives tractable prescriptions to describe a wide range of properties 
of strongly coupled field theories. For instance, a very neat and simple proposal was 
put forward in \cite{Calabrese:2004eu,Calabrese:2005zw}, to compute the entanglement entropy (EE) in field theories. In 
particular, the holographic description of quantum entanglement known as holographic EE (HEE)
have proven to be elegant in computing the EE of quantum field theories with conformal 
symmetry \cite{Ryu:2006bv,Ryu:2006ef}. The result for EE in two dimensional conformal field theories is well known,
 however the result in higher dimensions would be extremely difficult to compute. The 
 holographic prescription gives a handle to compute such quantities. The holographic
 calculation of EE starts with the Ryu-Takayanagi (RT) proposal which states that the
 HEE  of an asymptotically $AdS_{(d+1)}$ spacetime is equal to EE of a $d$- dimensional 
 $CFT$ living on the boundary of the gravitational theory in the bulk. The formula for the 
 HEE reads \cite{Ryu:2006bv,Ryu:2006ef}
 \begin{equation}
  S_{HEE}= \frac{Area (\gamma_A)}{4 G_{(d+1)}}
 \end{equation}
where $\gamma_A$ corresponds to the static minimal surface extending from the subsystem $A$, 
at the boundary of the asymptotically $AdS$ spacetime to the bulk. Thereafter, HEE 
calculation in various scenarios has been carried out extensively \cite{Bhattacharya:2012mi}-\cite{Saha:2018jjb}. 

It has also been realized that EE is useful in studying systems away from equilibrium. 
An important question that can be raised in this context is whether there exists
a relation analogous to the first law of thermodynamics. In \cite{Bhattacharya:2012mi,Allahbakhshi:2013rda}, the question 
was answered in the affirmative. It was found there that in the ultra-violet (UV) limit, 
the HEE gives a thermodynamics like relation. It was named as entanglement thermodynamics
and a notion of entanglement temperature came up with this observation. However, the 
entanglement temperature definition arising in the UV limit is quite different from
the well known thermodynamic temperature. This has lead to the investigation of finding 
a generalized temperature that agrees with the entanglement temperature in the UV limit and 
the Hawking temperature of the black hole in the infra-red (IR) limit \cite{Kim:2016jwu,Saha:2019ado}. 

Quantum complexity is another important quantity in the theory of quantum information. 
The quantity gives a measure of difficulty in performing a particular task. 
The proposal with which computations are usually carried out is the holographic
subregion complexity proposal (HSC) \cite{Mohsen}. It states that for a subsystem $A$ in the boundary, 
the HSC can be calculated from the formula 
\begin{equation}
 C_V=\frac{V(\gamma_A)}{8\pi R G_{(d+1)}}
\end{equation}
where $V(\gamma_A)$ is the maximal volume enclosed by the RT surface in the bulk, and
$R$ is the radius of curvature of the spacetime.

There are other proposals as well to compute the complexity holographically.
The proposal was put forward in \cite{sus1,sus2}. The prescription to obtain the complexity of a stste
is to calculate the volume of the Einstein-Rosen bridge (ERB) and is given by
\begin{equation}
C_V(t_L, t_R)=\frac{V(t_L, t_R)}{RG}
\end{equation}
where $V$ represents the spatial volume of the ERB. This volume is the maximum volume bounded by the CFT spatial slices at times $t_L$, $t_R$ on the two boundaries.

Then other proposal to calculate the holographic complexity is that of the bulk action 
computed on the Wheeler-De Witt patch \cite{Brown:2015lvg}
\begin{equation}
C_W=\frac{A(W)}{\pi \hbar}
\end{equation}
where $A(W)$ is the action cal;culated on the Wheeler-De Witt patch $W$.

The majority of the analyses of the quantities described above have been carried out for
systems which are relativistic in nature \cite{Ben-Ami:2016qex}-\cite{Karar:2019wjb}. Relatively few investigations has been done 
for non-relativistic systems. In \cite{Chakraborty:2014lfa}, the Lifshitz system in $3+1$- dimensions, which
is a well known non-relativistic system, was studied and thermodynamics like law for 
entanglement entropy of perturbed Lifshitz spacetime was obtained. In \cite{Karar:2017org}, the HSC of the
perturbed Lifshitz spacetime was computed, and a relation analogous to entanglement thermodynamics
was obtained in the context of HSC. 

In this paper, we set out to investigate the holographic quantities for a Lifshitz black hole \cite{Balasubramanian:2009rx}.
We first obtain the finite part of the HEE of the Lifshitz black hole, and then study its infrared (IR) and ultra-violet (UV) limits. We then proceed to investigate the near horizon behavior of the HEE to study the divergence structure. 
Next we obtain the change in HEE between a Lifshitz black hole and a pure Lifshitz spacetime. This we call the renormalized EE.
Then we proceed to write down a thermodynamic like relation involving the change in HEE
by introducing the notion of generalized temperature. The generalized temperature is defined 
in such a way that it gives the Hawking temperature of the Lifshitz black hole in the 
IR limit. In the UV limit, it gives rise to the entanglement temperature. 

We then look into other information theoretic quantities holographically.
We start by computing the HSC and then look at its near horizon limit.
Interestingly we see that the HSC has a logarithmic divergence in addition to the UV divergence.
The logarithmic divergence is absent in the Schwarzschild-$AdS$ case in $(3+1)$ - dimensions \cite{Roy:2017kha}.
The logarithmic divergence owes its origin to the power of the blackening factor
appearing in the spacetime metric. Then we compute the Fisher information metric holographically by
following the prescription in \cite{Lashkari:2015hha}. Then we compute the fidelity susceptibility from the 
fidelity expanded upto second order in the perturbation. We observe that the fidelity susceptibility can be 
related to the Fisher information metric upto a dimensional dependent constant. Finally, we compute the 
fidelity susceptibility by following the proposal in \cite{MIyaji:2015mia}. We observe that this does not match with
the Fisher information metric computed holographically.

The organization of the paper is as follows. The basic setup is discussed in section (\ref{2}).
This contains a short description of Lifshitz black hole metric and integrals related to
computation of HEE and HSC. We have computed the HEE for Lifshitz black hole in section (\ref{3}).
Firstly, the HEE has been computed analytically without any approximation then it has been
analyzed for IR and UV approximation. Then the near horizon behavior of HEE has been checked
in this section. The concept of generalized temperature has been introduced in section (\ref{4}).
In this section we have clearly shown that in IR regime the generalized temperature is equal to
the black hole temperature plus some correction terms. Those correction terms becomes negligible 
when the subsystem length becomes very large. The HSC has been computed in section (\ref{5}).
The holographic Fisher information metric and fidelity susceptibility has been discussed in section (\ref{6})


\section{Lifshitz black hole}\label{2}
We are interested in computing HEE and HC for Lifshitz black hole which asymptotically approaches
to Lifshitz spacetime in near boundary limit. For Lifshitz black hole one may consider
the following $3+1$ - dimensional action \cite{Balasubramanian:2009rx}

\begin{equation}
S=  \frac{1}{2}\int d^4x ~ (R-2\varLambda) -\int d^4x \left(e^{-2 \Phi} \frac{F^2}{4} +\frac{m^2 }{2}A^2+
(e^{-2\Phi -1})\right).
\end{equation}
A solution to this action is given by
\begin{eqnarray}
 ds^2 = -f(r)\frac{dt^2}{r^{2z}}+\frac{dx^2 + dy^2}{r^2}+\frac{dr^2}{r^2 f(r)} {\label{metric}}\\
 \Phi=-\frac{1}{2} \log(1+\frac{r^2}{r_h^2})~; \hspace{3mm} A=\frac{f(r)}{r^2}dt
\end{eqnarray}
with
\begin{equation}{\label{lapse}}
 f(r)=1-\frac{r^2}{r_h^2}
\end{equation}
where $r_h$ is the horizon radius of the black hole. 
This solution enjoys an anisotropic scaling with the space
and time scaling as $(x,y) \rightarrow (\lambda x, \lambda y)$
and $t \rightarrow \lambda^z t$. Here $z$ is called the dynamical exponent.
Such theories are non relativistic 
as they do not obey Lorentz invariance and have a lot of 
importance in the study of condensed matter systems
near the quantum critical point. The spacetime metric \eqref{metric} reduces to that of pure Lifshitz 
spacetime (vacuum solution) in the near boundary limit ($r \rightarrow 0$).

\noindent The Hawking temperature and entropy of the 
black hole are given by
\begin{equation}
 T_{h}=\frac{1}{2\pi r_h^2}~; ~~~~~~ S_h = \frac{lL}{4G r_h^2}~.
\end{equation}

\noindent We now choose the shape of the subsystem  to be strip like
for computing the HEE and subregion HC. 
The strip like subsystem lies at 
the boundary of the Lifshitz black hole with the specifications $-\frac{l}{2}\leq x\leq \frac{l}{2}$;
$0\leq y \leq L$. According to the prescription in \cite{Ryu:2006bv,Ryu:2006ef} the HEE is proportional
to the static minimal area of the hypersurface in the bulk whose
boundary coincides with the boundary of the
subsystem at $r=0$. To evaluate that minimal area we parametrize the hypersurface as $r= r(x)$ and 
leave the $y$-direction independent. 
With this parametrization we get the area of the hypersurface to be
\begin{equation}
 A=L\int_{-l/2}^{l/2} dx\;\; \frac{1}{r^2}\sqrt{1+\frac{r'(x)^2}{f(r)}} 
\end{equation}
where $r'(x)\equiv \frac{dr(x)}{dx}$. 
Using the standard procedure of minimization, we obtain the minimal surface 
specified by the following equation
\begin{equation}
 \frac{dr(x)}{dx}=\sqrt{f(r)\left(\frac{r_t^4}{r^4}-1 \right)}
\end{equation}
where $r_t$ is the turning point of the minimal surface. 
This condition for minimal surface can be used to
get the minimal surface area and the subsystem length as
\begin{equation}
 A=2L\int_{r_c}^{r_t}dr\; \frac{1}{r^2 \sqrt{f(r) (1-r^4/r_t^4)}}~;~~~~ 
 l=2\int_0^{r_t} dr\; \frac{(r/r_t)^2}{\sqrt{f(r) (1-r^4/r_t^4)}}  
\end{equation}
where $r_c$ is the UV cutoff. The minimal volume under the same hypersurface is given by
\begin{eqnarray}
 V &=& 2L \int_{r_c}^{r_t} dr\; \frac{1}{r^3 \sqrt{f(r)}}\; x(r) \nonumber\\
 &=& 2L \int_{r_c}^{r_t} dr\; \frac{1}{r^3 \sqrt{f(r)}}\; \int_r^{r_t} ds \frac{(s/r_t)^2}{\sqrt{f(s) (1-s^4/r_t^4)}}~.
\end{eqnarray}
For the sake of simplicity we choose a new coordinate $u=r/r_t$. With this change of variables,  
the lapse function takes the form $f(u)=1-u^2/u_0^2$, where $u_0=r_h/r_t$ and 
the scaled UV cutoff $\delta$ is defined as
 $\delta=r_c/r_t$. Therefore, the expressions for the subsystem length, minimal hypersurface area, 
minimal volume takes the form
\begin{equation}\label{len}
 l=2r_t \int_0^1 du\; \frac{u^2}{\sqrt{(1-u^4)f(u)}} 
\end{equation}

\begin{equation}\label{area}
 A=\frac{2L}{r_t}\int_{\delta}^{1}du \; \frac{1}{u^2\sqrt{(1-u^4)f(u)}} 
\end{equation}

\begin{equation}\label{vol}
V= \frac{2L}{r_t}\int_{\delta}^{1}du \; \frac{1}{u^3\sqrt{f(u)}} \int_u^1 ds\;\frac{s^2}{\sqrt{(1-s^4)f(s)}}~.
\end{equation}
In the rest of the paper we have used the above expressions for computing different 
holographic quantities.  



\section{Holographic Entanglement Entropy}\label{3}

The integrals involved in the expressions for subsystem length \eqref{len} and
area \eqref{area} integrals contains a term $1/\sqrt{f(u)}$. 
When $f(u)=1$, the background geometry reduces to that of $(3+1)$-dimensional
pure Lifshitz spacetime \cite{Balasubramanian:2009rx,Kachru:2008yh}. The computation of HEE for such a system
is easy due to the absence of $1/\sqrt{f(u)}$ term. 
In the case of the Lifshitz black hole, the integrals become non-trivial, 
but can be done analytically. For that we have to expand $1/\sqrt{f(u)}$ binomially as
\begin{equation}\label{exp}
 \frac{1}{\sqrt{f(u)}}=\sum_{n=0}^{\infty} \frac{\Gamma (n+\frac{1}{2})}{\sqrt{\pi}\Gamma (n+1)}u^n~.
\end{equation}
Using the above expression we obtain the subsystem length from eq.\eqref{len}, which reads
\begin{equation}\label{len1}
 \frac{l}{r_t}=\sum_{n=0}^{\infty}\frac{\Gamma (n+\frac{1}{2})\Gamma(\frac{n}{2}+\frac{3}{4})}{2\Gamma(n+1)\Gamma(\frac{n}{2}+\frac{5}{4})}
 \left( \frac{r_t}{r_h} \right)^{2n}~.
\end{equation}
When the subsystem length $l$ is small (i.e, $l/r_h \ll 1$) or the Hawking temperature
of the black hole is small, we have $r_t \ll r_h$. 
The above sum can then be terminated for some value of $n$
as the higher order terms can be neglected. In $r_t\rightarrow r_h$ limit,  
one cannot terminate the series unlike the previous case. We need to check the behavior
of $l$ for large values of $n$. The expression \eqref{len1}
goes as $\sim \frac{1}{n}\left(\frac{r_t}{r_h}\right)^{2n}$ for large values of $n$. 
The expression for the subsystem length given in eq.\eqref{len1} can be written as
\begin{equation}
 \frac{l}{r_t} = \frac{\sqrt{\pi}\Gamma(3/4)}{2\Gamma(5/4)}
 +\sum_{n=1}^{\infty} \frac{\Gamma (n+\frac{1}{2})\Gamma(\frac{n}{2}+\frac{3}{4})}{2\Gamma(n+1)\Gamma(\frac{n}{2}+\frac{5}{4})}
 \left( \frac{r_t}{r_h} \right)^{2n}.
\end{equation}
For large values of $n$ the second term goes as $\sim \frac{1}{2\sqrt{2}\;n}\; $.
Therefore the comparison test for infinite series implies that the 
series is divergent in $r_h \rightarrow r_t$ limit.
We separate the divergence part to rewrite the above expression as
\begin{equation}
  \frac{l}{r_t} = \frac{\sqrt{\pi}\Gamma(3/4)}{2\Gamma(5/4)}
  +\sum_{n=1}^{\infty} \frac{\Gamma (n+\frac{1}{2})\Gamma(\frac{n}{2}+\frac{3}{4})}{2\Gamma(n+1)\Gamma(\frac{n}{2}+\frac{5}{4})}
 \left(1-\frac{1}{\sqrt{2}n} \right)\left( \frac{r_t}{r_h} \right)^{2n} -\frac{1}{\sqrt{2}} \log\left(1-r_t^2/r_h^2 \right)~.
\end{equation}
In the above expression, the divergent piece $\frac{1}{\sqrt{2}} \log\left(1-r_t^2/r_h^2 \right)$
has been separated out. Now since the hypersurface cannot penetrate the 
black hole horizon \cite{Hubeny:2012ry}, we use the relation 
Now using the approximation $r_t \simeq r_h(1-\epsilon)$, where $\epsilon$ 
is very small, and obtain 
\begin{equation}
 \frac{l}{r_h}=k_1 -\frac{1}{\sqrt{2}}\log(2 \epsilon)+ \mathcal{O}(\epsilon)
\end{equation}
with
\begin{equation}
 k_1= \frac{\sqrt{\pi}\Gamma(3/4)}{2\Gamma(5/4)}
  +\sum_{n=1}^{\infty} \frac{\Gamma (n+\frac{1}{2})\Gamma(\frac{n}{2}+\frac{3}{4})}{2\Gamma(n+1)\Gamma(\frac{n}{2}+\frac{5}{4})}
 \left(1-\frac{1}{\sqrt{2}n} \right).
\end{equation}
We now proceed to compute the area \eqref{area}
using the same expansion of the lapse factor (eq. \ref{exp}). This gives 
\begin{equation}
 A=\frac{2L}{r_t}\sum_{n=0}^{\infty} \frac{\Gamma (n+\frac{1}{2})}{\sqrt{\pi}\Gamma (n+1) u_0^{2n}}
 \int_{\delta}^{1} du\; \frac{u^{2n-2}}{\sqrt{1-u^4}}~.
\end{equation}
Looking at the above expression one can see that the integral is divergent
for $n=0$ in the $\delta \rightarrow 0$ limit. 
Performing the computation for the $n=0$ term separately, we get
\begin{equation}\label{A0}
 A_{n=0}=\frac{2L}{r_t}\left(\frac{1}{\delta} -\frac{\sqrt{\pi} \Gamma (3/4)}{\Gamma(1/4)} \right).
\end{equation}
We observe that the first term in the above expression is divergent.
After computing the $A_{n\geq 1}$ terms, we combine them to get the total area as
\begin{equation}\label{area1}
 A=\frac{2L}{r_t}\left(\frac{1}{\delta} -\frac{\sqrt{\pi} \Gamma (3/4)}{\Gamma(1/4)}+
 \sum_{n=1}^{\infty} \frac{\Gamma (n+\frac{1}{2})\Gamma(\frac{n}{2}-\frac{1}{4})}
 {4\Gamma (n+1)\Gamma(\frac{n}{2}+\frac{1}{4})}\left(1-\frac{4 \Gamma(\frac{n}{2}+\frac{1}{4}) \delta^{2n-1}}
 {\sqrt{\pi}\Gamma(\frac{n}{2}-\frac{1}{4})} \right)\left( \frac{r_t}{r_h} \right)^{2n}\right).
\end{equation}
The above expression shows that area has an UV divergence going as $\sim \frac{1}{\delta}$.
This UV divergence is exactly similar to that of the $(3+1)$ - dimensional $AdS$ black brane.
So this divergence is universal irrespective of the underlying theory being 
relativistic or non relativistic.
Further, $A_{n=0}$ is the minimal area of the hypersurface for the 
pure Lifshitz spacetime, which expectedly has a UV divergent term.
From eq.(s) (\ref{A0}, \ref{area1}), we can now obtain the finite 
part of the minimal area of the hypersurface to be 
\begin{equation}
 A_{finite} =\frac{2L}{r_t}\left( -\frac{\sqrt{\pi} \Gamma (3/4)}{\Gamma(1/4)}+
 \sum_{n=1}^{\infty} \frac{\Gamma (n+\frac{1}{2})\Gamma(\frac{n}{2}-\frac{1}{4})}
 {4\Gamma (n+1)\Gamma(\frac{n}{2}+\frac{1}{4})}\left( \frac{r_t}{r_h} \right)^{2n}\right).
\end{equation}
where we have taken the $\delta \rightarrow 0$ limit and also subtracted 
the divergent term proportional to $1/\delta$.

\noindent We now use the gamma function identity $\Gamma(p+1)=p \;\Gamma(p)$ to rewrite the above result as
\begin{eqnarray}
  A_{finite} &=& \frac{2L}{r_t}\left( -\frac{\sqrt{\pi} \Gamma (3/4)}{\Gamma(1/4)}+
 \sum_{n=1}^{\infty}\frac{1}{4}\left(1+\frac{2}{2n-1} \right) 
 \frac{\Gamma (n+\frac{1}{2})\Gamma(\frac{n}{2}+\frac{3}{4})}
 {\Gamma (n+1)\Gamma(\frac{n}{2}+\frac{5}{4})}\left( \frac{r_t}{r_h} \right)^{2n}\right)\nonumber\\
 &=& \frac{2L}{r_t}\left( -\frac{2\sqrt{\pi} \Gamma (3/4)}{\Gamma(1/4)}+\frac{l}{2 r_t}+
 \sum_{n=1}^{\infty}\frac{1}{2(2n-1)}  
 \frac{\Gamma (n+\frac{1}{2})\Gamma(\frac{n}{2}+\frac{3}{4})}
 {\Gamma (n+1)\Gamma(\frac{n}{2}+\frac{5}{4})}\left( \frac{r_t}{r_h} \right)^{2n}\right)
\end{eqnarray}
where in the second line of the equality we have used the expression for subsystem length \eqref{len1}. 
The third term in the above expression, for large values of $n$  goes as 
$\sim \frac{1}{2\sqrt{2} n^2}\left( \frac{r_t}{r_h} \right)^{2n}$. Using this fact,
$A_{finite}$ can be recast as 
\begin{eqnarray}
 A_{finite}&=&\frac{2L}{r_t}\left( -\frac{2\sqrt{\pi} \Gamma (3/4)}{\Gamma(1/4)}+\frac{l}{2 r_t}+
 \sum_{n=1}^{\infty}\left(\frac{1}{2(2n-1)}  
 \frac{\Gamma (n+\frac{1}{2})\Gamma(\frac{n}{2}+\frac{3}{4})}
 {\Gamma (n+1)\Gamma(\frac{n}{2}+\frac{5}{4})}-\frac{1}{2\sqrt{2}n^2}\right)
 \left( \frac{r_t}{r_h} \right)^{2n} \right.\nonumber \\ 
 && \hspace{5cm}\left. +\sum_{n=1}^{\infty} \frac{1}{2\sqrt{2}n^2}\left( \frac{r_t}{r_h} \right)^{2n} \right)\nonumber\\
 &=& \frac{2L}{r_t}\left( -\frac{2\sqrt{\pi} \Gamma (3/4)}{\Gamma(1/4)}+\frac{l}{ 2 r_t}+
 \sum_{n=1}^{\infty}\left(\frac{1}{2(2n-1)}  
 \frac{\Gamma (n+\frac{1}{2})\Gamma(\frac{n}{2}+\frac{3}{4})}
 {\Gamma (n+1)\Gamma(\frac{n}{2}+\frac{5}{4})}-\frac{1}{2\sqrt{2}n^2}\right)
 \left( \frac{r_t}{r_h} \right)^{2n} \right.\nonumber \\ 
 && \hspace{5cm}\left. + \frac{1}{2\sqrt{2}}Li_{2}\left[\left(\frac{r_t}{r_h}\right)^2\right] \right)~.
\end{eqnarray}
This leads to the finite EE
\begin{eqnarray}
S_{finite}&=& \frac{A_{finite}}{4G_4} \nonumber\\
&=& \frac{L}{2G_4 r_t}\left( -\frac{2\sqrt{\pi} \Gamma (3/4)}{\Gamma(1/4)}+\frac{l}{ 2 r_t}+
\sum_{n=1}^{\infty}\left(\frac{1}{2(2n-1)}  
\frac{\Gamma (n+\frac{1}{2})\Gamma(\frac{n}{2}+\frac{3}{4})}
{\Gamma (n+1)\Gamma(\frac{n}{2}+\frac{5}{4})}-\frac{1}{2\sqrt{2}n^2}\right)
\left( \frac{r_t}{r_h} \right)^{2n} \right.\nonumber \\ 
&& \hspace{5cm}\left. + \frac{1}{2\sqrt{2}}Li_{2}\left[\left(\frac{r_t}{r_h}\right)^2\right] \right)~.
\end{eqnarray}
Now using the definition of HEE and the approximation $r_t=r_h(1-\epsilon)$ (IR limit), we have
\begin{eqnarray}\label{s1}
 S_{finite}^{(IR)} &=& S_h + \frac{L}{2G_4 r_h} \left(k_2 + k_3 \; \epsilon +k_4 \; \epsilon \log \epsilon \right)\nonumber\\
 &\simeq& S_h +\frac{L}{2G_4 r_h} k_2+ \mathcal{O}(\epsilon)
\end{eqnarray}
where 
\begin{equation}
 S_h= \frac{lL}{4G_4 r_h^2}
\end{equation}
is the Bekenstein--Hawking entropy of the Lifshitz black hole and
\begin{eqnarray}
 k_2&=& -\frac{2\sqrt{\pi} \Gamma (3/4)}{\Gamma(1/4)}+ \sum_{n=1}^{\infty}\left(\frac{1}{2(2n-1)}  
 \frac{\Gamma (n+\frac{1}{2})\Gamma(\frac{n}{2}+\frac{3}{4})}
 {\Gamma (n+1)\Gamma(\frac{n}{2}+\frac{5}{4})}-\frac{1}{2\sqrt{2}n^2}\right) +\frac{1}{2\sqrt{2}}\xi(2), \nonumber \\
k_3 &=& \frac{2( \log 2 -1)}{2\sqrt{2}}~;~~~~~~~ k_4= \frac{1}{\sqrt{2}}~.
 \end{eqnarray}
Thus, we find that the holographic entanglement entropy in the 
IR limit is the thermal entropy plus correction terms. 
Since the black hole temperature goes as $T_{h} \sim 1/r_h^2$,
so in terms of the black hole temperature the finite part of HEE reads
\begin{equation}\label{t1}
 S_A^{finite} \sim T_h\left(1 +c_1 \frac{1}{\sqrt{T_h}}\right)
\end{equation}
with $c_1$ being some numerical constant.

\noindent If the subsystem length $l$ is small $\left(l/r_h <<1\right)$ then the 
bulk extension will be near the boundary. Therefore the turning point $r_t$ of the
RT surface will be far away from the black hole horizon $r_h$. In this approximation
$r_t/r_h \ll 1$ we can take only first few terms of binomial expansion of $1/\sqrt{f(u)}$
in the expression \eqref{exp}. This approximation may be called the UV limit. In UV limit
the expression for finite part of HEE is given by
\begin{eqnarray}
 S_{finite}^{(UV)}&=& \frac{L}{4G_4 l}\left(-4\pi \left(\frac{\Gamma(3/4)}{\Gamma(1/4)} \right)^2 
 +\frac{l^2}{r_h^2}\frac{1}{12}\left(\frac{\Gamma(1/4)}{\Gamma(3/4)}\right)^2  \right. \nonumber \\
 &&\left.~~~~~~~~~~~~+ \frac{l^4}{r_h^4}\frac{3}{16\pi}\left(\frac{\Gamma(1/4)}{\Gamma(3/4)}\right)^2\left(\frac{1}{5}-\frac{1}{432}\left(\frac{\Gamma(1/4)}{\Gamma(3/4)}\right)^2\right) \right).
\end{eqnarray}
We would like to point out that the  expressions for 
subsystem length \eqref{len1} and hypersurface area
\eqref{area1} are exact in the 
sense that no approximation has been made. 
We are now proceed to see the behavior of the UV cutoff dependent divergences in the 
near horizon approximation.
The near horizon approximation is important when we deal with large
subsystem length ($l/r_h >>1$) or high temperature
black holes cases. In both the cases the horizon approaches the turning point of 
the hypersurface ($r_h \rightarrow r_t$). This approximation therefore implies $u_0 \sim 1$. Hence the
integrals in eq.(s) (\ref{len},\ref{area}) receives most of the contribution when $u \sim 1$. 
We therefore make a Taylor expansion of the lapse function around $u \sim u_0$ to get
\begin{eqnarray}
f(u)&=& f(u_0)+(u-u_0)f'(u_0)+\frac{(u-u_0)^2}{2}f''(u_0) \cdots \nonumber \\
  &\approx& 2\left(1-\frac{u}{u_0} \right),
\end{eqnarray}
where we have neglected the higher order terms as $u-u_0\ll 1$.
With this approximation the subsystem length is as follows
\begin{equation}\label{lenh}
\frac{l}{r_t}=\sum_{n=0}^{\infty}\frac{\Gamma(n+\frac{1}{2}) \Gamma(\frac{n+3}{4})}
{2\sqrt{2}\;\Gamma(n+1)\Gamma(\frac{n+5}{4})} \left(\frac{r_t}{r_h} \right)^{n}.
\end{equation}
The area integral under this approximation is given by
\begin{equation}
 A=\frac{L}{r_t}\sum_{n=0}^{\infty}\frac{\sqrt{2}\Gamma(n+\frac{1}{2})}{\sqrt{\pi}\Gamma(n+1)}\frac{1}{u_0^n}
 \int_{\delta}^{1} du\; \frac{u^{n-2}}{\sqrt{1-u^4}}~.
\end{equation}
The above expression contains integrals which are divergent for $n=0,1$.
We compute them separately now. These reads
\begin{eqnarray}
 A_{n=0}&=&\frac{\sqrt{2}L}{r_t}\left(\frac{1}{\delta} -\frac{\sqrt{\pi} \Gamma (3/4)}{\Gamma(1/4)} \right) \nonumber\\
 A_{n=1}&=& \frac{L}{\sqrt{2}r_h} \left[\int_{\delta}^{1}du\; \frac{1}{u} 
 + \sum_{m=1}^{\infty}\frac{\Gamma (m+\frac{1}{2})}{\sqrt{\pi}\Gamma (m+1)}\int_{\delta}^{1}du\; u^{4m-1} \right] \nonumber\\
 &=& \frac{L}{\sqrt{2}r_h}\left(- \log \delta +\frac{\log 4}{4} \right)
\end{eqnarray}
where we have used the expansion
\begin{equation}
 1/\sqrt{1-u^4} = \sum_{n=0}^{\infty} \frac{\Gamma (n+\frac{1}{2})}{\sqrt{\pi}\Gamma (n+1)}u^{4n}
\end{equation}
in the computation of the $A_{n-1}$ integral. Now computing $A_{n\geq 2}$ terms
and gathering all terms, we write the expression for area in the near horizon
approximation to be
\begin{equation}\label{areah}
 A= \frac{\sqrt{2}L}{r_t} \left( \frac{1}{\delta}-\frac{\sqrt{\pi} \Gamma (3/4)}{\Gamma(1/4)} 
 + \frac{r_t}{2r_h}\left(- \log \delta +\frac{\log 4}{4}\right)
 +\sum_{n=2}^{\infty}\frac{\Gamma(n+\frac{1}{2})\Gamma(\frac{n-1}{4})}{4 \Gamma(n+1)\Gamma(\frac{n+1}{4})}
 \left(\frac{r_t}{r_h} \right)^n\right).
\end{equation}
Since we are in the near horizon approximation $r_h \rightarrow r_t$,
we use the expression \eqref{lenh} to recast the expression for area as
\begin{equation}
 A \sim \frac{L}{r_h} \left(\frac{c_2}{\delta}+ c_3 \log \delta +c_4 \frac{l}{r_h}+c_5 \right)
\end{equation}
where $c_2, c_3, c_4, c_5$ are numerical constants. Interestingly
we observe that in near horizon approximation the expression for area
contains logarithmic divergence term in addition to the usual $1/\delta$ divergence term.
Furthermore, the finite part of the area 
have similar form as that obtained in the IR limit. Hence the finite part of HEE would have
similar temperature dependence as in eq.\eqref{t1}.

\subsection{Generalized temperature}
\begin{figure}[ht!]
\centering
\begin{tabular}{cccc}
\includegraphics[width=0.45\textwidth]{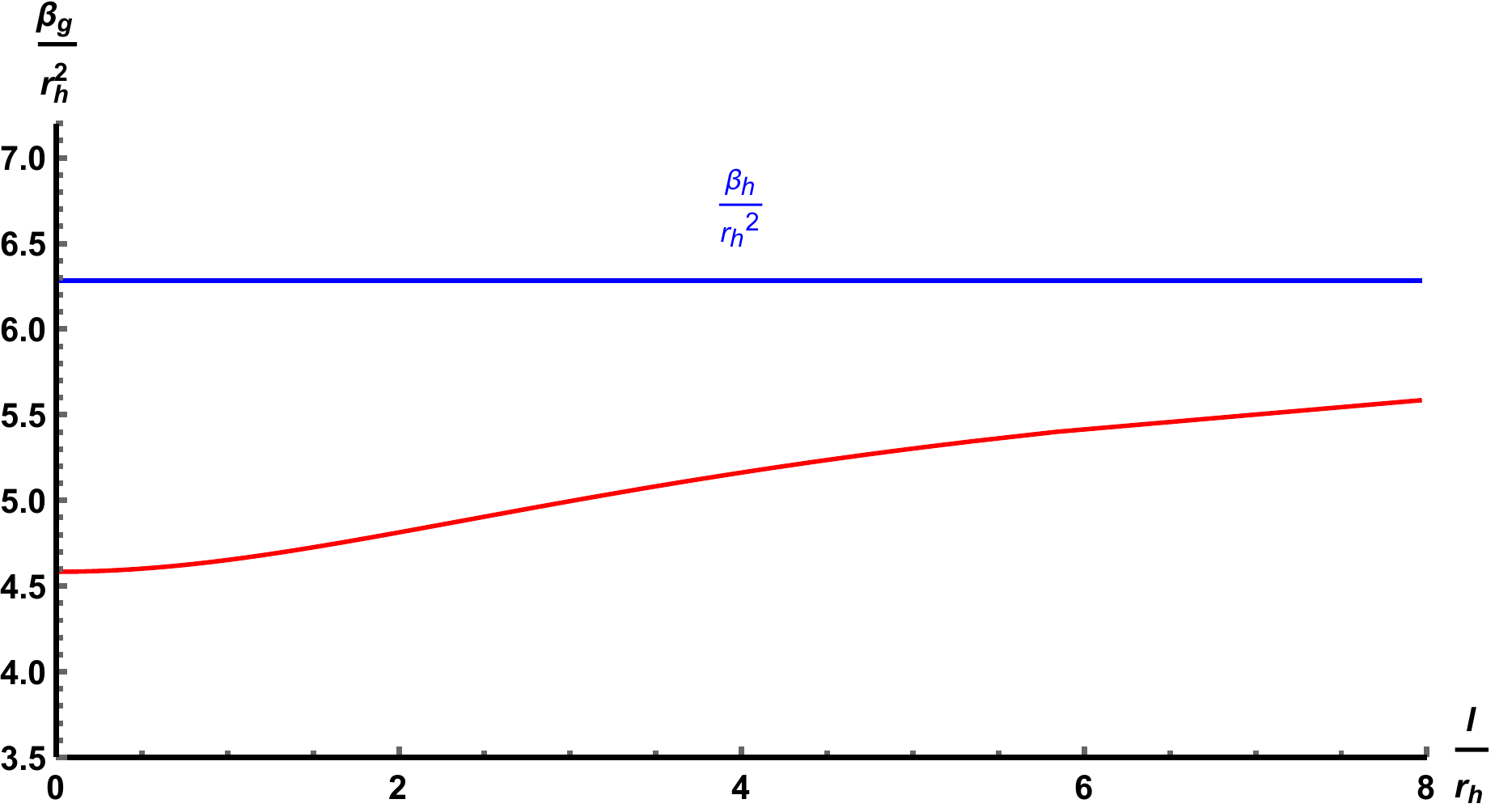}  &
\includegraphics[width=0.45\textwidth]{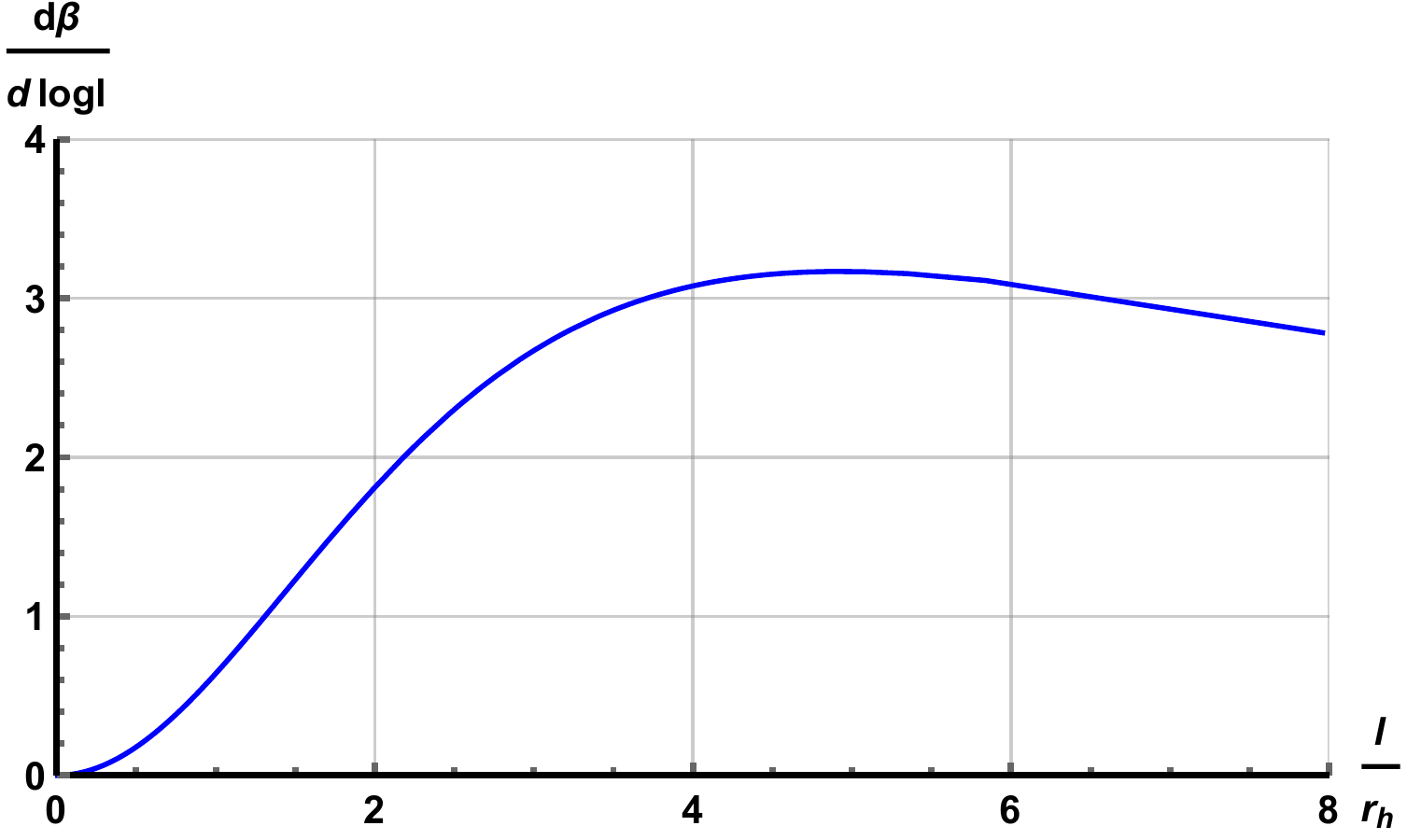} \\
\textbf{(a)}  & \textbf{(b)}  \\[6pt]
\end{tabular}
\caption{\footnotesize \textbf{(a)} Variation of $\frac{\beta_g}{r_h^2}=\frac{1}{T_g r_h^2}$ with subsystem length,
\textbf{(b)} Renormalization Group flow (The $\frac{d \beta}{d \log l}$ axis has been scaled five times for 
visualization purpose)}
\label{RG}
\end{figure}

{\label{4}}
The Lifshitz black hole as described in section (\ref{2}) satisfies the first law of black hole thermodynamics \cite{Balasubramanian:2009rx}
\begin{equation}
dE =T_h \; dS_h
\end{equation}
 leading to 
 \begin{equation}
 E=\int T_h \; dS_h = T_h S_h/2
 \end{equation}
  where the energy $E$ is given by 
\begin{equation}{\label{energy}}
E=\frac{lL}{16 \pi G r_h^4}.
\end{equation}
The relation is important since
it relates the near boundary quantity ($E$) to the near horizon quantities ($S_h, T_h$). 
It can be seen from eq.\eqref{s1} that in the IR limit ($r_t \rightarrow r_h$) the leading contribution to
the HEE comes from the thermal entropy $S_h$. As we depart from the IR limit, the HEE gets quantum corrections
due to microscopic properties of the underlying quantum system.
Keeping this point in mind we may ask for a quantity 
called the generalized temperature ($T_g$) which is different from the black hole temperature $T_h$.
 We define the generalized temperature in the following way
\begin{equation}\label{genT}
 \frac{1}{T_g}= \frac{S_{REE}}{E};~~~~ S_{REE}=S_A - S_A^{(0)}=\frac{1}{4G_4}\left(A-A^{(0)} \right)
\end{equation}
where $S_{REE}$ is the HEE of the Lifshitz black hole and $S_A^{(0)}$ is the  HEE of the pure Lifshitz spacetime. The 
hypersurface area and subsystem length as obtained for pure Lifshitz spacetime are given by \cite{Karar:2017org}
\begin{equation}\label{pure}
 A_0=\frac{2L}{\delta r_t^{(0)}}- \frac{4 \pi L }{ l}\left(\frac{\Gamma(3/4)}{\Gamma(1/4)}\right)^2; ~~~~
 \frac{l}{r_t^{(0)}}=\frac{\sqrt{\pi}\Gamma(3/4)}{2 \; \Gamma(5/4)}
\end{equation}
Using this and eq.\eqref{energy} the generalized temperature reads
\begin{equation}{\label{genT1}}
 \frac{1}{T_g}=\frac{2\pi r_h^2}{l}\left(\frac{4 \pi  }{ l}\left(\frac{\Gamma(3/4)}{\Gamma(1/4)}\right)^2 
 -\frac{2\sqrt{\pi} \Gamma (3/4)}{r_t \; \Gamma(1/4)}
+\frac{1}{2r_t} \sum_{n=1}^{\infty} \frac{\Gamma (n+\frac{1}{2})\Gamma(\frac{n}{2}-\frac{1}{4})}
 {4\Gamma (n+1)\Gamma(\frac{n}{2}+\frac{1}{4})}\left(\frac{r_t}{r_h} \right)^{2n}\right).
\end{equation}
From the above expression we see that $T_g$ is a function of $l$ and $r_t$, but $r_t$ itself depends on the subsystem
length $l$. So we conclude that $T_g$ is a function of $l$ alone. Therefore the generalized temperature
depends on the subsystem size. We now discuss the behavior of the generalized temperature in extreme limits.
In the IR limit ($r_t \rightarrow r_h$), the generalized temperature takes the form 
\begin{equation}{\label{genTir}}
 \frac{1}{T_g}= \frac{1}{T_h} + \frac{r_h^3}{l} \alpha +\frac{r_h^4}{l^2}\beta 
 \end{equation}
with \begin{eqnarray}
 \alpha &=& 2\pi \left(-\frac{4\sqrt{\pi}\Gamma(3/4)}{\Gamma(1/4)}+
 \sum_{n=1}^{\infty} \frac{\Gamma (n+\frac{1}{2}) \Gamma(\frac{n}{2}+\frac{3}{4})}
 {(2n-1)\Gamma(n+1)\Gamma(\frac{n}{2}+\frac{5}{4})}\right) \nonumber \\
 \beta &=& 8 \pi^2\left(\frac{\Gamma(3/4)}{\Gamma(1/4)}\right)^2.
\end{eqnarray}
It is evident from eq.\eqref{genTir} that the generalized temperature yields the thermodynamic temperature $T_h$
in the large subsystem size limit ($l/r_h \gg1$). The sub-leading terms are due to 
quantum entanglement. 
In the UV limit ($r_t/r_h \ll 1$), the generalized temperature is given by
\begin{equation}{\label{genTuv}}
 \frac{1}{T_g}=\frac{\pi r_h^2}{6}\left(\frac{\Gamma(1/4)}{\Gamma(3/4)}\right)^2 \left(1 +\frac{l^2}{r_h^2}\gamma\right)
\end{equation}
where
\begin{equation}
 \gamma =\frac{9}{20 \pi } -\frac{1}{192 \pi}
 \left(\frac{\Gamma(1/4)}{\Gamma(3/4)}\right)^4~.
\end{equation}
In the $l/r_h \rightarrow 0$ limit (that is UV limit), the generalized temperature temperature attains a constant value. 
It is completely a non-relativistic phenomena (for relativistic systems the generalized temperature approaches [REF]
to zero in the UV limit . This is one of the central results of our paper. 
In the UV limit, the Lifshitz blackhole asymptotically approaches to pure Lifshitz spacetime
which is a vacuum solution resulting to zero temperature. So it is possible to get surprised  
after looking into our result. But we should not confuse the actual temperature of the background geometry
with the generalized temperature. The defining form of generalized temperature \eqref{genT}
is ratio of two quantities, $S_A - S_A^{(0)}$ and $E$. Both of these terms approaches to zero
when $l \rightarrow 0$, resulting into ($\frac{0}{0}$) form for $T_g$. So our result is
mathematically consistent. This non zero generalized temperature($T_g$) in 
$l \rightarrow 0$ limit can be called ``entanglement temperature''($T_{ent}$).
The origin of entanglement temperature is solely microscopic (quantum entanglement)
and has nothing to do with the macroscopic properties of the system.
The expression for the entanglement temperature follows from eq.\eqref{genTuv} and is given by 
\begin{equation}
 T_{ent}= \frac{6}{\pi r_h^2}\left(\frac{\Gamma(3/4)}{\Gamma(1/4)}\right)^2.
\end{equation}
Figure \ref{RG}.(a) and our analysis in the IR regime, it is evident that the 
generalized temperature ($T_g$)
becomes the Hawking temperature ($T_h$) for large subsystem length. 
To characterize the thermal and quantum nature of the system
we study the flow of $T_g$. We 
study the flow of the generalized temperature by studying the variation of $\frac{d \beta}{d \log l}$ 
with $l$. This has been shown in figure \ref{RG}.(b). From the flow, we observe that it has a maximum near 
a critical value $\frac{l_c}{r_h}=4.91$. Above this value of the subsystem size, the system behave as a thermal system
and below this value the system behaves as a quantum system.

\section{Holographic subregion complexity} \label{5}
The subregion HC is proportional to the volume under the minimal hypersurface whose boundary coincides with the 
the boundary of the subsystem lying at the boundary at a fixed time. Using the approximation \eqref{exp}
in the expression for volume \eqref{vol}, we get
\begin{equation}
 V=\frac{L}{r_t}\sum_{n=0}^{\infty} \sum_{m=0}^{\infty} \left(V_1 +V_2+V_3 \right)
 \left(\frac{r_t}{r_h} \right)^{2n+2m}
\end{equation}
where
\begin{eqnarray}\label{terms}
 V_1&=& \frac{2\Gamma(n+\frac{1}{2})\Gamma(m+\frac{1}{2})}{\pi\;(2n+3)(2m+2n+1) \Gamma(n+1) \Gamma(m+1)} 
 \delta^{2m+2n+1} \nonumber \\
 V_2 &=& -\frac{\Gamma(n+\frac{1}{2})\Gamma(m+\frac{1}{2})}
 {4 \sqrt{\pi}(m-1)\Gamma(n+1)\Gamma(m+1)\Gamma\left(\frac{2n+5}{4} \right)} \delta^{2m-2} \nonumber \\
 V_3 &=& \frac{\Gamma(n+\frac{1}{2})\Gamma(m+\frac{1}{2})\Gamma\left(\frac{2m+2n+1}{4}\right)}
 {4 \sqrt{\pi}(m-1)\Gamma(n+1)\Gamma(m+1)\Gamma\left(\frac{2m+2n+3}{4}\right)}~.
\end{eqnarray}
Now we should checkout for the divergences in the above mentioned terms as
the expression for volume contains double sum which extends from $m, n =0$ to $\infty$.
Due to the presence of the term $\left(\frac{r_t}{r_h} \right)^{2n+2m}$, we expect
the divergence to occur near $r_h \rightarrow r_t$. We start analyzing the $V_1$
term. For large values of $(m, n)$,  $V_1$ varies as
\begin{equation}
 V_1 \sim \frac{1}{n^{\frac{3}{2}} m^{\frac{1}{2}}(n+m)} < \frac{1}{n^{\frac{3}{2}} m^{\frac{3}{2}}}~.
\end{equation}
We can now argue that
$\sum_{n,m=0}^{\infty} V_1\left(\frac{r_t}{r_h} \right)^{2n+2m}$ 
is convergent. If $p(m,n)$ and $q(m,n)$ are two sequences of real numbers
with $m$ and $n$ being positive integers, 
then the series $\sum_{n,m=0}^{\infty}p(m,n)$ is convergent
if for each values of $(m,n)$ we have
$|p(m,n)| \leq |q(m,n)|$.
If we choose $q(m,n)=\frac{1}{n^{\frac{3}{2}} m^{\frac{3}{2}}}$, then 
$\sum_{n,m=0}^{\infty}\frac{1}{n^{\frac{3}{2}} m^{\frac{3}{2}}}=1/(\xi(3/2))^2 $,
 where $\xi(3/2)$ is the Reimann Zeta function. Therefore the double sum over $V_1$ term
converges absolutely. Moreover it is multiplied by $\delta^{2m+2n+1}$, which tends to zero.
So we can neglect the contribution from $V_1$. 
Let us now look at the $V_2$ term which has a UV cutoff dependent term $\delta^{2m-2}$.
For $m\geq 2$, this term will have negligible contribution but for $m=0, 1$, this term is UV divergent. To get the exact form of UV divergence 
we compute the volume for $m=0$ and $1$ separately.
It is expected that the $m=1$ term should give a logarithmic UV divergence.
The volume integral for $m=0$ and $m=1$ are given by 
\begin{eqnarray}
 V_{m=0} &=&\frac{L}{r_t}\sum_{n=0}^{\infty}\left( \frac{r_t}{r_h}\right)^{2n}
 \left(\frac{\Gamma(n+\frac{1}{2})\Gamma(\frac{3}{4}+\frac{n}{2})}{4\Gamma(n+1)\Gamma(\frac{5}{4}+\frac{n}{2})}\frac{1}{\delta^2}
  -\frac{\Gamma(n+\frac{1}{2})}{\sqrt{\pi}(2n+3)\Gamma(n+1)}\delta^{2n+1} \right. \nonumber \\
  &&\hspace{5cm} -\left. \frac{\Gamma(n+\frac{1}{2})\Gamma(\frac{1}{4}+\frac{n}{2})}{4 \Gamma(n+1)
  \Gamma(\frac{3}{4}+\frac{n}{2})}\right) \nonumber \\
 V_{m=1} &=& \frac{L}{r_t}\sum_{n=0}^{\infty}\left( \frac{r_t}{r_h}\right)^{2n}
 \left(- \frac{\Gamma(n+\frac{1}{2})\Gamma(\frac{3}{4}+\frac{n}{2})}{4\Gamma(n+1)\Gamma(\frac{5}{4}+\frac{n}{2})} \log \delta
 +\frac{\Gamma(n+\frac{1}{2})}{\sqrt{\pi}(2n+3)\Gamma(n+1)}\delta^{2n+3}\log \delta  \right. \nonumber\\
 && \hspace{3cm}\left. +\frac{\Gamma(n+\frac{1}{2})\Gamma(\frac{3}{4}+\frac{n}{2})}{16\Gamma(n+1)\Gamma(\frac{5}{4}+\frac{n}{2})}
 \left(H_n\left[\frac{n}{2}-\frac{1}{4} \right]-H_n\left[\frac{n}{2}+\frac{1}{4} \right] \right) \right)~.
\end{eqnarray}
In the $r_h \rightarrow r_t$ limit, we use the expression for subsystem length \eqref{len1} to get
\begin{equation}
 V_{m=0} \sim \frac{lL}{(r_h)^2}\left(\frac{b_1}{\delta^2}+b_4 \right);~~~ V_{m=1}\sim
 \frac{lL}{(r_h)^2}\left( b_2\; \log \delta + b_5 \right)
\end{equation}
where $b_1, b_2, b_3$ and $b_4$ are numerical constants. Finally we look at the $V_3$ term. 
The analysis for $V_3$ term is more complicated.
The detailed analysis is given in Appendix (\ref{appB}). 
This term in $r_h \rightarrow r_t$ limit varies as
\begin{equation}
 V_3 \sim \frac{l}{r_h}\left( b_3 \frac{r_h}{l} + b_6 \right)~.
\end{equation}
Combining all these terms we get the volume to be
\begin{equation}
 V \sim \frac{lL}{(r_h)^2}\left(\frac{b_1}{\delta^2}+b_2\; \log \delta +b_3 \frac{r_h}{l} + b_7 \right)
\end{equation}
where $b_7$ is another numerical constant. Therefore in terms of the black hole temperature, 
the expression for the finite part of holographic subregion complexity varies as
\begin{equation}\label{t2}
 C_{finite} \sim  T_h\left(b_7  +b_3\frac{1}{ \sqrt{T_h}}\right).
\end{equation}
Therefore, the holographic entanglement entropy \eqref{t1} and holographic subregion complexity \eqref{t2} have the same kind of 
temperature dependence.
We now proceed to evaluate the volume integral in the near horizon limit. In this limit the volume becomes
\begin{equation}
 V=\frac{L}{r_t}\sum_{n=0}^{\infty} \sum_{m=0}^{\infty} \left(V_1 +V_2+V_3 \right)
 \left(\frac{r_t}{r_h} \right)^{n+m}
\end{equation}
where
\begin{eqnarray}
 V_1&=& \frac{\Gamma(n+\frac{1}{2})\Gamma(m+\frac{1}{2})}{\pi\;(n+3)(m+n+1) \Gamma(n+1) \Gamma(m+1)} 
 \delta^{m+n+1} \nonumber \\
 V_2 &=& -\frac{\Gamma(n+\frac{1}{2})\Gamma(m+\frac{1}{2})}
 {4 \sqrt{\pi}(m-2)\Gamma(n+1)\Gamma(m+1)\Gamma\left(\frac{n+5}{4} \right)} \delta^{m-2} \nonumber \\
 V_3 &=& \frac{\Gamma(n+\frac{1}{2})\Gamma(m+\frac{1}{2})\Gamma\left(\frac{m+n+1}{4}\right)}
 {4 \sqrt{\pi}(m-2)\Gamma(n+1)\Gamma(m+1)\Gamma\left(\frac{m+n+3}{4}\right) }~.
\end{eqnarray}
The terms $V_1$ and $V_3$ have almost the same pattern as in the
previous case. So we can say that the term $V_1$ is convergent and have negligible contribution 
due to the UV cutoff $\delta^{m+n+1}$ and the term $V_3$ goes as
\begin{equation}
V_3 \sim \frac{lL}{(r_h)^2}\left( d_5 \frac{r_h}{l} + d_9 \right)~.
\end{equation}
Now the term $V_2$ is divergent for $m \leq 2$, so we compute
the volume term for $m=0, 1, 2$ separately (see Appendix (\ref{appC})).
In the limit $r_h \rightarrow r_t$, they behave as
\begin{eqnarray}
V_{m=0} \sim \frac{lL}{(r_h)^2} \left(d_1\frac{1}{\delta^2}+ d_6\right) \nonumber \\
V_{m=1} \sim \frac{lL}{(r_h)^2} \left(d_2\frac{1}{\delta}+ d_7 \right) \nonumber \\
V_{m=2} \sim  \frac{lL}{(r_h)^2} \left(d_3 \log \delta+ d_8 \right)~.
\end{eqnarray}
Combining all the volume terms we get
\begin{equation}
 V \sim \frac{lL}{(r_h)^2} \left(d_1\frac{1}{\delta^2}+d_2\frac{1}{\delta} + 
 d_3 \log \delta+ d_{10} +d_5 \frac{r_h}{l}\right)~.
\end{equation}
So the finite part of the holographic subregion complexity varies with the black hole temperature as
\begin{equation}
 V_{finite}\sim T_h\left(d_{10}  +b_9 \frac{1}{\sqrt{T_h}}\right).
\end{equation}

\section{Fisher information metric and Fidelity susceptibility}{\label{6}}
In this section, we shall compute the Fisher information metric and the fidelity susceptibility
for the Lifshitz black hole using holographic prescriptions
\cite{Lashkari:2015hha,MIyaji:2015mia,Banerjee:2017qti}. 
In the context of quantum information theory there exists two well notions of
distance between two quantum states. They are the Fisher information metric \cite{paris2009quantum} and the fidelity susceptibility \cite{Bures}
(or called the Bures metric).
From the literature \cite{Banerjee:2017qti}, the definition of the Fisher information metric is given by
\begin{equation}
G_{F,\lambda \lambda}=\langle \delta \rho\; \delta \rho\rangle_{\lambda \lambda}^{(\sigma)}=
\frac{1}{2}tr \left(\delta \rho \frac{d}{d(\delta \lambda)} \log(\sigma + \delta \lambda \delta \rho)|_{\delta \lambda=0} \right)
\end{equation}
where $\delta \rho$ is a small deviation from the density matrix $\sigma$.

\noindent On the other hand the fidelity susceptibility is given by
\begin{equation}\label{fidelity}
G_{\lambda \lambda }= \partial^2_{\lambda} F ;~~~~F= tr\sqrt{\sqrt{\sigma_{\lambda}} \rho_{\lambda + \delta \lambda} \sqrt{\sigma_{\lambda}}}
\end{equation} 
where $\rho$ and $\sigma $ are the final and initial density matrices, $F$ is called the fidelity between the two states.

\noindent The holographic computation of the Fisher information metric from relative EE 
was put forward in \cite{Lashkari:2015hha}. The Fisher information metric is given by 
\begin{equation}{\label{fisher}}
G_{F,mm}=\frac{\partial^2 }{\partial m^2}S_{rel}(\rho_m \parallel \rho_0);~~~~~~~ S_{rel}(\rho_m \parallel \rho_0)=\Delta \langle H_{\rho_0} \rangle-\Delta S
\end{equation}
where $m$ is a perturbation parameter, $\Delta S$ is the change in entanglement entropy from the
vacuum state and $\Delta \langle H_{\rho_0} \rangle$ is the change in modular Hamiltonian.
With this basic background in place we first compute the Fisher information metric for the 
Lifshitz black hole.
We consider that the background is slightly perturbed from the pure Lifshitz spacetime but the
subsystem length $l$ fixed. Then the inverse of the lapse function \ref{lapse} can be written as
\begin{equation}
\frac{1}{f(r)}=\frac{1}{1-{r^2\over r_h^2}}=1+m r^2+ m^2 r^4.
\end{equation}
where $m=\frac{1}{r_h^2}$, is the perturbation parameter in the bulk. As the underlying geometry 
has been changed from Lifshitz spacetime to asymptotic Lifshitz spacetime and we have not changed the subsystem
length the turning point of the bulk extension will change as
\begin{equation}{\label{rt2}}
 r_t=r_t^{(0)}+m~r_t^{(1)}+m^2~r_t^{(2)} 
\end{equation}
where $r_t^{(0)}$ is the turning point for pure Lifshitz spacetime and $r_t^{(1)}$, $r_t^{(2)}$
are first and second order corrections to the turning point.On the other hand the subsystem length $l$
can be obtained from (\ref{len},\ref{len1}) upto second order in perturbation as
\begin{equation}{\label{len2}}
 \frac{l}{r_t}=a_0 +m~a_1 r_t^2 +m^2~ a_2 r_t^4 
\end{equation}
with
\begin{equation}
 a_0 =\frac{\sqrt{\pi}\Gamma\left(\frac{3}{4}\right)}{2~\Gamma\left(\frac{5}{4} \right)};~
 a_1 = \frac{\Gamma\left(\frac{3}{2}\right)\Gamma\left(\frac{5}{4}\right)}{2~\Gamma\left(\frac{7}{4} \right)};~
 a_2=\frac{\Gamma\left(\frac{5}{2}\right)\Gamma\left(\frac{7}{4}\right)}{4~\Gamma\left(\frac{9}{4} \right)}.
\end{equation}
Using the fact that the subsystem length is unchanged we obtain from equations (\ref{rt2},\ref{len2}), 
expressions for $r_t^{(0)}$, $r_t^{(1)}$ and $r_t^{(2)}$ as given below
\begin{equation}
 \frac{l}{r_t^{(0)}} =a_0,~~ r_t^{(1)}=-\frac{a_1}{a_0}(r_t^{(0)})^3,~~
 r_t^{(2)}=\left(3\frac{a_1^2}{a_0^2}-\frac{a_2}{a_0}\right)(r_t^{(0)})^5~~.
\end{equation}
The expression for extremal area of the bulk extension upto second order in perturbation parameter can be 
obtained from eq.(\ref{area}, \ref{area1}) as
\begin{equation}\label{area2}
 A=\frac{2L}{r_c}-a_0 \frac{L}{r_t}+ 3a_1Lr_t m+\frac{5}{3}a_2 Lr_t^3 m^2.
\end{equation}
As we are interested in computing the change in area due to a slight change in background we
use \eqref{rt2} to recast the above expression for area in the following form
\begin{equation}
 A=A^{(0)}+m~A^{(1)}+m^2~A^{(2)}
\end{equation}
where $A^{(0)}$, $A^{(1)}$ and $A^{(2)}$ are area for pure Lifshitz spacetime, first order and second order
corrections to the area for change in background. They have the following expressions
\begin{equation}
 A^{(0)} = \frac{2L}{r_c}-a_0 \frac{L}{r_t^{(0)}},~~ A^{(1)} =2a_1 Lr_t^{(0)},~~
 A^{(2)}=\left(\frac{2}{3}a_2 -\frac{a_1^2}{a_0} \right)L(r_t^{(0)})^3.
\end{equation}
It has been shown in \cite{Lashkari:2015hha}
that at first order in $m$ the relative entropy vanishes and in second order in $m$ the
relative entropy is given by $S_{rel}=-\Delta S$. Hence,
\begin{equation}
S_{rel}= -m^2\frac{A^{(2)}}{4G} =m^2 \frac{L l^3}{4G}\left(\frac{3a_1^2 -2a_0 a_2}{3a_0^4} \right).
\end{equation}
From eq. \eqref{fisher}, the Fisher information metric therefore reads
\begin{equation}\label{fisher0}
G_{F,mm}=  \frac{L l^3}{2G}\left(\frac{3a_1^2 -2a_0 a_2}{3a_0^4} \right).~.
\end{equation} 

\noindent In \cite{Banerjee:2017qti}, a proposal for computing the above quantity was given.
The proposal is to consider the difference  of two volumes yielding a finite expression
\begin{equation}
 \mathcal{F}=C_d(V-V_{(0)})
\end{equation}
where $ V$ is evaluated for a second order perturbation around pure Lifshitz spacetime 
. $C_d$ is a dimensionless constant which 
cannot be fixed from the first principles of the gravity side. We shall now apply this proposal to compute
the Fisher information metric for the Lifshitz black hole.
The change in volume under Ryu-Takayanagi minimal surface at second order in perturbation takes the form
\begin{equation}
V=V^{(0)}+m V^{(1)}+m^2 V^{(2)} 
\end{equation}
with
\begin{equation}
 V^{(0)}=\frac{L l}{2 r_c^2} -\frac{a_0^2 L}{2l}, ~~V^{(1)}= - mL l \left(\frac{a_1}{2a_0}+\frac{6-\pi}{8}\right),~~
  V^{(2)} = m^2 \frac{Ll^3}{a_0^3}\left(\frac{a_1^2}{a_0}-\frac{a_2}{2}+\frac{13-3\pi}{12}a_1 \right).
\end{equation}
The holographic dual of Fisher information metric is now defined as
\begin{equation}{\label{deb}}
G_{F,mm}=\partial_m^2 \mathcal{F};~~~~~~\mathcal{F}= C_d  \left(V-V_{(0)} \right)
\end{equation}
with the constant $C_d$ to be determined by requiring that the holographic dual Fisher information
metric from the above equation must agree with that obtained from the relative
entropy \eqref{fisher0}. The constant $C_d$ is therefore given by
\begin{equation}{\label{cd0}}
C_d=\frac{3a_1^2 -2a_0 a_2}{G(12a_1^2 -6a_0a_2 +(13-3\pi)a_0a_1)} ~.
\end{equation} 
We now look forward to compute the fidelity susceptibility holographically.
If one assumes that the states depend on a single parameter $\lambda$ then for pure states the 
fidelity \eqref{fidelity} reduces to 
\begin{equation}
\langle \Psi (\lambda) | \Psi (\lambda+ \delta \lambda) \rangle =1-G_{\lambda \lambda}  (\delta \lambda)^2+ \cdots~.
\end{equation}
The above expression immediately suggests that $G_{\lambda \lambda}$ is a 
measure of distance between two quantum states called the fidelity susceptibility . 
The holographic prescription for evaluating the fidelity susceptibility in
($d+1$) - dimensional $AdS$ spacetime is given by \cite{MIyaji:2015mia}
\begin{gather}{\label{fs}}
 G_{\lambda \lambda} = n_{d-1} \frac{Vol(\Sigma_{max})}{R^{d}}
\end{gather}
where $\Sigma_{max}$ is the maximum volume in the bulk that ends at the boundary of the bulk
at a fixed time slice. $R$ is the radius of curvature of $AdS$ spacetime and
$n_{d-1}$ is a $\mathcal{O}(1)$ constant.

\noindent For our case the fidelity susceptibility reads
\begin{eqnarray}\label{fs1}
 G_{\lambda \lambda}&=& n_{2} L l \int_{r_c}^{r_h} dr \frac{1}{r^3 \sqrt{1-{r^2\over r_h^2}}} \notag\\
 &=& \frac{n_{2} Ll}{r_h^2}\left(\frac{r_h^2}{2r_c^2}+B(-1,\frac{1}{2}) \right).
\end{eqnarray}
We see that the above expression for the fidelity susceptibility does not agree with the Fisher information
metric obtained in eq.\eqref{fisher0}.


\section{Conclusion}
In this paper, we have computed different information theoretic quantities holographically in the context of a non-relativistic ($3+1$)-dimensional Lifshitz black hole. Our main focus has been the following. To begin with we have looked at the
Ryu-Takayanagi area which is related to the holographic entanglement entropy and second is the minimal volume under the Ryu-takayanagi area
which is related  to holographic subregion complexity. The finite part of the holographic entanglement entropy approaches the black hole entropy
in the infrared ($r_t \rightarrow r_h$) regime. This clearly depicts that the  entanglement entropy becomes thermal entropy in the
high temperature limit even in the non-relativistic background. In the ultra-violet ($r_t\ll r_h$) limit, the finite
part of holographic entanglement entropy goes as $S_{finite}^{(UV)} \sim \frac{1}{l}\left(constant + \mathcal{O}(l^2)\right)$. This result departs with respect to the subleading terms from the relativistic counterpart comprising the $(3+1)$-dimensional $SAdS$ black hole where 
$S_{finite}^{(UV)} \sim \frac{1}{l}\left(constant + \mathcal{O}(l^3) \right)$.
Further, the holographic entanglement entropy has a logarithmic divergence in addition to the usual $1/\delta$ divergence in the near horizon approximation.
We have then introduced the notion of a generalized temperature in terms of the renormalized holographic entanglement entropy. 
The variation of the generalized temperature with the subsystem length $l$ shows that the generalized temperature
reduces to the black hole temperature, that is the Hawking temperature, at large subsystem length (infrared limit). This therefore implies that
our choice of the definition for the generalized temperature is correct. It has been observed that the generalized temperature leads to a thermodynamics like
law $E=T_g S_{REE}$. It is also interesting to note that
the generalized temperature has a constant value in the ultraviolet limit ($l \rightarrow0$). 
This is a new result which does not have any counterpart in the relativistic background, namely, the $(3+1)$-dimensional $SAdS$ black hole. 
We have then observed departures from relativistic results in case of the holographic subregion complexity. Both in the $3+1$-dimensional relativistic and non-relativistic
cases, the holographic subregion complexity suffers $1/\delta^2$ divergences, but in the non-relativistic case
we have $\log \delta$ divergence which was absent in the relativistic case. The near horizon approximation have the same type of divergences in both the cases.
The holographic Fisher information metric has been computed next from the concept of relative entropy \cite{Lashkari:2015hha}. We have also computed the holographic subregion complexity in the ulta-violet limit upto second order
in the perturbation parameter. Using the proposal in \cite{Banerjee:2017qti}, we have used this result to obtain an 
expression for the Fisher information metric upto an undetermined constant. We have equated this result of the
Fisher information metric with that obtained from the relative entropy to determine the undetermined constant. The constant is found to be a number.
The holographic fidelity susceptibility has also been computed. We find that
there is a mismatch between the expressions for the Fisher information metric and the holographic fidelity susceptibility. These two quantities are related in the context of quantum information. 
A possible reason for this mismatch may be the following. In case of the Fisher information metric, an integration up to the
turning point of the Ryu-takayanagi surface is performed. However, in case of the  fidelity susceptibility, an integration
up to the horizon radius of the black is involved.
A similar observation has already been made in the relativistic background in \cite{Karar:2019wjb}.


\section*{Acknowledgment}
 SG acknowledges the support of IUCAA, Pune for the Visiting Associateship programme.

\section*{Appendix} 
 
 \begin{appendix}

\section{Analysis of $V_3$ term of volume}\label{appB}
The $V_3$ term as given in equation \eqref{terms} for large $(m,n)$ goes as
\begin{equation}
 V_3 \sim \frac{1}{m^{3/2}\sqrt{n(m+n)}}.
\end{equation}
So the sum over $n$ is divergent by comparison test. We perform the sum over $m$ to get
\begin{eqnarray}
 V_3&=&\sum_{n=0}^{\infty} \sum_{m=0}^{\infty}\left(\frac{r_t}{r_h} \right)^{2n+2m}\frac{\Gamma(n+\frac{1}{2})\Gamma(m+\frac{1}{2})\Gamma\left(\frac{2m+2n+1}{4}\right)}
 {4 \sqrt{\pi}(m-1)\Gamma(n+1)\Gamma(m+1)\Gamma\left(\frac{2m+2n+3}{4}\right) } \nonumber \\
 &=& \sum_{n=0}^{\infty} \left(\frac{r_t}{r_h} \right)^{2n}\frac{\Gamma(n+\frac{1}{2})}{128\; \Gamma(n+1)}
 \left(\frac{5\Gamma(\frac{2n+7}{4})}{\Gamma(\frac{2n+9}{4})}\; HPFQR\left[\{1,1,\frac{7}{4},\frac{9}{4},\frac{9}{4}+\frac{n}{2}\},\{2,2,\frac{5}{2},\frac{9}{4}+\frac{n}{2}\},1\right]\right. \nonumber\\
 && \left. \hspace{3cm}- \frac{32\Gamma(\frac{2n+1}{4})}{\Gamma(\frac{2n+3}{4})}\; HPFQR\left[\{-\frac{1}{2},\frac{1}{2},\frac{3}{4},\frac{1}{4}+\frac{n}{2}\},\{\frac{1}{2},\frac{1}{2},\frac{3}{4}+\frac{n}{2}\},1\right]\right)
\end{eqnarray}
where $HPFQR$ is the $HypergeometricPFQRegularized$ function. For large values of $n$ the summation term
varies as $ \sim \frac{1}{n}\left(\frac{r_t}{r_h} \right)^{2n} $. On the other hand the expression for 
length $\left(\frac{l}{r_t}\right)$ as presented in \eqref{len1} also varies as  $ \sim \frac{1}{n}\left(\frac{r_t}{r_h} \right)^{2n} $
in large $n$ limit. Therefore in $r_h \rightarrow r_t$ limit the $V_3$ term varies as
\begin{equation}
 V_3 \sim \left(a \frac{l}{r_h}+b \right)
\end{equation}
where $a, b$ are numerical constants.

\section{Exact computation of Volume for $m=0, 1, 2$ in near horizon limit }\label{appC}
\begin{eqnarray}
 V_{m=0} &=& \frac{L}{r_t}\sum_{n=0}^{\infty}\left(\frac{r_t}{r_h} \right)^{n} \frac{\Gamma(n+\frac{1}{2})}{\sqrt{\pi}\Gamma(n+1)}
 \left(\int_0^1 du\; \frac{u^{n+2}}{\sqrt{1-u^4}} \int_\delta^1 du\; \frac{1}{u^3}- \int_\delta^1 du\; \frac{1}{u^3}\int_0^u ds\frac{s^{n+2}}{\sqrt{1-u^4}}\right) \nonumber \\
 &=&\frac{L}{r_t}\left(\sum_{n=0}^{\infty} \left(\frac{r_t}{r_h} \right)^{n}\frac{\Gamma(n+\frac{1}{2}) \Gamma(\frac{n+3}{4})}
{8\;\Gamma(n+1)\Gamma(\frac{n+5}{4})}\frac{1}{\delta^2}-\sum_{n=0}^{\infty} \left(\frac{r_t}{r_h} \right)^{n} 
\frac{\Gamma(n+\frac{1}{2}) \Gamma(\frac{n+1}{4})}
{8\;\Gamma(n+1)\Gamma(\frac{n+3}{4})}\right) \nonumber \\
&=&\frac{L}{r_t}\left(\frac{l}{2\sqrt{2}\;r_t}\frac{1}{\delta^2}- \sum_{n=0}^{\infty} \left(\frac{r_t}{r_h} \right)^{n} 
\frac{\Gamma(n+\frac{1}{2}) \Gamma(\frac{n+1}{4})}
{8\;\Gamma(n+1)\Gamma(\frac{n+3}{4})} \right),
\end{eqnarray}
where in the last line we have used the expression for subsystem length in near horizon limit \eqref{lenh}.
\begin{eqnarray}
  V_{m=1} &=& \frac{L}{r_t}\sum_{n=0}^{\infty}\left(\frac{r_t}{r_h} \right)^{n+1} \frac{\Gamma(n+\frac{1}{2})}{2\sqrt{\pi}\Gamma(n+1)}
 \left(\int_0^1 du\; \frac{u^{n+2}}{\sqrt{1-u^4}} \int_\delta^1 du\; \frac{1}{u^2}- \int_\delta^1 du\; \frac{1}{u^2}\int_0^u ds\frac{s^{n+2}}{\sqrt{1-u^4}}\right) \nonumber \\
 &=&\frac{L}{r_t}\left(\sum_{n=0}^{\infty} \left(\frac{r_t}{r_h} \right)^{n+1}\frac{\Gamma(n+\frac{1}{2}) \Gamma(\frac{n+3}{4})}
{8\;\Gamma(n+1)\Gamma(\frac{n+5}{4})}\frac{1}{\delta}-\sum_{n=0}^{\infty} \left(\frac{r_t}{r_h} \right)^{n} 
\frac{\Gamma(n+\frac{1}{2}) \Gamma(\frac{n+2}{4})}
{8\;\Gamma(n+1)\Gamma(\frac{n+4}{4})}\right) \nonumber \\
&=&\frac{L}{r_t}\left(\frac{l}{2\sqrt{2}\;r_t}\frac{1}{\delta}- \sum_{n=0}^{\infty} \left(\frac{r_t}{r_h} \right)^{n} 
 \frac{\Gamma(n+\frac{1}{2}) \Gamma(\frac{n+2}{4})}
{8\;\Gamma(n+1)\Gamma(\frac{n+4}{4})}\right)
\end{eqnarray}
and 
\begin{eqnarray}
  V_{m=2} &=& \frac{L}{r_t}\sum_{n=0}^{\infty}\left(\frac{r_t}{r_h} \right)^{n+2} \frac{3\;\Gamma(n+\frac{1}{2})}{8\sqrt{\pi}\Gamma(n+1)}
 \left(\int_0^1 du\; \frac{u^{n+2}}{\sqrt{1-u^4}} \int_\delta^1 du\; \frac{1}{u}- \int_\delta^1 du\; \frac{1}{u}\int_0^u ds\frac{s^{n+2}}{\sqrt{1-u^4}}\right) \nonumber \\
 &=&\frac{L}{r_t}\sum_{n=0}^{\infty} \left(\frac{r_t}{r_h} \right)^{n+2} \frac{3\;\Gamma(n+\frac{1}{2})}{8\sqrt{\pi}\Gamma(n+1)}
 \left( - \frac{\sqrt{\pi} \Gamma(\frac{n+3}{4})}{4\Gamma(\frac{n+5}{4})}\log \delta
 +\frac{\delta^{n+3}\log \delta}{n+3} \right. \nonumber\\
 && \left. \hspace{5cm} +\frac{\sqrt{\pi}\Gamma(\frac{n+3}{4})}{16 \Gamma(\frac{n+5}{4})}\left(H_n(\frac{n-1}{4})-H_n(\frac{n+1}{4}) \right)\right) \nonumber \\
 &\approx& \frac{L}{r_t} \left(-\frac{3\sqrt{2\; }l}{16\; r_t}\log \delta 
 +\sum_{n=0}^{\infty} \left(\frac{r_t}{r_h} \right)^{n+2}\frac{3\;\Gamma(n+\frac{1}{2})\Gamma(\frac{n+3}{4})}{128\Gamma(n+1)\Gamma(\frac{n+5}{4})}\left(H_n(\frac{n-1}{4})-H_n(\frac{n+1}{4}) \right) \right)
\end{eqnarray}
\end{appendix}


\begin{thebibliography}{99}

\bibitem{Witten:1998qj}
  E.~Witten,
  Adv.\ Theor.\ Math.\ Phys.\  {\bf 2} (1998) 253
  
\bibitem{Maldacena:1997re}
  J.~M.~Maldacena,
  Int.\ J.\ Theor.\ Phys.\  {\bf 38} (1999) 1113
   [Adv.\ Theor.\ Math.\ Phys.\  {\bf 2} (1998) 231]
   
\bibitem{Aharony:1999ti}
  O.~Aharony, S.~S.~Gubser, J.~M.~Maldacena, H.~Ooguri and Y.~Oz,
  Phys.\ Rept.\  {\bf 323} (2000) 183
  
\bibitem{Calabrese:2004eu} 
  P.~Calabrese, J.~L.~Cardy,
  ``Entanglement entropy and quantum field theory'',
  J.\ Stat.\ Mech.\  0406  (2004) P06002.
  
\bibitem{Calabrese:2005zw} 
  P.~Calabrese, J.~L.~Cardy,
  ``Entanglement entropy and quantum field theory: A Non-technical 
introduction'',
  Int.\ J.\ Quant.\ Inf.\  4 (2006) 429.
  
\bibitem{Vidal:2002rm}
  G.~Vidal, J.~I.~Latorre, E.~Rico and A.~Kitaev,
  ``Entanglement in quantum critical phenomena,''
  Phys.\ Rev.\ Lett.\  {\bf 90} (2003) 227902
  
\bibitem{Kitaev:2005dm}
  A.~Kitaev and J.~Preskill,
  ``Topological entanglement entropy,''
  Phys.\ Rev.\ Lett.\  {\bf 96} (2006) 110404
  
  \bibitem{Wen:2006}
  M.~Levin and X.~G.~Wen,
  ``Detecting topological order in a ground state wave function'',
  Phys.\ Rev.\ Lett.\  {\bf 96} (2006) 110405
  
\bibitem{Hartnoll:2009sz}
  S.~A.~Hartnoll,
  ``Lectures on holographic methods for condensed matter physics,''
  Class.\ Quant.\ Grav.\  {\bf 26} (2009) 224002
  
\bibitem{Cai:2012sk}
  R.~G.~Cai, S.~He, L.~Li and Y.~L.~Zhang,
  ``Holographic Entanglement Entropy in Insulator/Superconductor Transition,''
  JHEP {\bf 1207} (2012) 088

  
\bibitem{CasalderreySolana:2011us}
  J.~Casalderrey-Solana, H.~Liu, D.~Mateos, K.~Rajagopal and U.~A.~Wiedemann,
  ``Gauge/String Duality, Hot QCD and Heavy Ion Collisions,''
  book:Gauge/String Duality, Hot QCD and Heavy Ion Collisions. Cambridge, UK: Cambridge University Press, 2014
  [arXiv:1101.0618 [hep-th]].   

\bibitem{Ryu:2006bv} 
  S.~Ryu, T.~Takayanagi,
  ``Holographic derivation of entanglement entropy from AdS/CFT'',
  Phys.\ Rev.\ Lett.\  96 (2006) 181602.
  
\bibitem{Ryu:2006ef} 
  S.~Ryu, T.~Takayanagi,
  ``Aspects of Holographic Entanglement Entropy'',
  JHEP 0608 (2006) 045.

 \bibitem{Bhattacharya:2012mi}
     J.~Bhattacharya, M.~Nozaki, T.~Takayanagi, T.~Ugajin,
     Phys.\  Rev.\  Lett.\  110 (2013) 091602.

 \bibitem{Allahbakhshi:2013rda}
 D.~Allahbakhshi, M.~Alishahiha, A.~Naseh,
  JHEP 1308 (2013) 102.

\bibitem{Solodukhin:2006xv}
  S.~N.~Solodukhin,
  Phys.\ Rev.\ Lett.\  {\bf 97} (2006) 201601  

\bibitem{Hubeny:2012ry}
  V.~E.~Hubeny,
  JHEP {\bf 1207} (2012) 093
  doi:10.1007/JHEP07(2012)093   
  
\bibitem{Chaturvedi:2016kbk}
  P.~Chaturvedi, V.~Malvimat and G.~Sengupta,
  Phys.\ Rev.\ D {\bf 94} (2016) no.6,  066004
  
\bibitem{Karar:2018ecr}
  S.~Karar, D.~Ghorai and S.~Gangopadhyay,
  Nucl.\ Phys.\ B {\bf 938} (2019) 363

\bibitem{Saha:2018jjb}
  A.~Saha, S.~Karar and S.~Gangopadhyay,
  Eur.\ Phys.\ J.\ Plus {\bf 135} (2020) no.2,  132  
  
\bibitem{Kim:2016jwu}
  K.~S.~Kim and C.~Park,
  Phys.\ Rev.\ D {\bf 95} (2017) no.10,  106007
  doi:10.1103/PhysRevD.95.106007  
 
\bibitem{Saha:2019ado}
  A.~Saha, S.~Gangopadhyay and J.~P.~Saha,
  Phys.\ Rev.\ D {\bf 100} (2019) no.10,  106008
  doi:10.1103/PhysRevD.100.106008
\bibitem{Mohsen} M.~Alishahiha, 
     Phys. Rev. D 92 (2015) 126009. 
     
\bibitem{sus1} L.~Susskind, 
Fortsch.Phys. 64 (2016) 24, Addendum: Fortsch.Phys. 64 (2016) 44-48.
     
     
\bibitem{sus2}  D.~Stanford, L.~Susskind, 
 Phys. Rev. D 90 (2014) 126007.      
     
\bibitem{Brown:2015lvg}
  A.R.~Brown, D.A.~Roberts, L.~Susskind, B.~Swingle, Y.~Zhao,
  Phys.\ Rev.\ D {\bf 93} (2016) no.8,  086006. 
  


\bibitem{Ben-Ami:2016qex}
  O.~Ben-Ami and D.~Carmi,
  JHEP {\bf 1611} (2016) 129
  
\bibitem{Couch:2016exn}
  J.~Couch, W.~Fischler, P.~H.~Nguyen,
  JHEP {\bf 1703} (2017) 119.
  
\bibitem{Chapman:2016hwi}
  S.~Chapman, H.~Marrochio and R.~C.~Myers,
  JHEP {\bf 1701} (2017) 062.
  
\bibitem{Carmi:2016wjl}
  D.~Carmi, R.~C.~Myers, P.~Rath,
  JHEP {\bf 1703} (2017) 118.  
  
 
  
 \bibitem{Carmi:2017jqz}
  D.~Carmi, S.~Chapman, H.~Marrochio, R.~C.~Myers and S.~Sugishita,
  JHEP {\bf 1711} (2017) 188.
 
 \bibitem{Caputa:2017yrh}
  P.~Caputa, N.~Kundu, M.~Miyaji, T.~Takayanagi and K.~Watanabe,
  JHEP {\bf 1711} (2017) 097.
  
  \bibitem{Agon:2018zso}
  C.~A.~Agón, M.~Headrick, B.~Swingle,
  JHEP {\bf 1902} (2019) 145.
  
\bibitem{Chapman:2018dem}
  S.~Chapman, H.~Marrochio and R.~C.~Myers,
  JHEP {\bf 1806} (2018) 046.
  
\bibitem{Chapman:2018lsv}
  S.~Chapman, H.~Marrochio, R.C.~Myers,
  JHEP {\bf 1806} (2018) 114.  

\bibitem{Karar:2018hvy}
  S.~Karar, S.~Gangopadhyay and A.~S.~Majumdar,
  Int.\ J.\ Mod.\ Phys.\ A {\bf 34} (2019) no.01,  1950003  
  
\bibitem{Ghosh:2019jgd}
  A.~Ghosh and R.~Mishra,
  arXiv:1907.11757 [hep-th].
 
\bibitem{Karar:2019wjb}
  S.~Karar, R.~Mishra and S.~Gangopadhyay,
  Phys.\ Rev.\ D {\bf 100} (2019) no.2,  026006
  doi:10.1103/PhysRevD.100.026006
  
\bibitem{Chakraborty:2014lfa}
  S.~Chakraborty, P.~Dey, S.~Karar, S.~Roy,
  JHEP {\bf 1504} (2015) 133.  
  
\bibitem{Karar:2017org}
  S.~Karar, S.~Gangopadhyay,
  Phys.\ Rev.\ D {\bf 98} (2018) no.2,  026029.  
  
\bibitem{Balasubramanian:2009rx}
  K.~Balasubramanian and J.~McGreevy,
  Phys.\ Rev.\ D {\bf 80} (2009) 104039
  doi:10.1103/PhysRevD.80.104039  
  
\bibitem{Roy:2017kha}
  P.~Roy and T.~Sarkar,
  Phys.\ Rev.\ D {\bf 96} (2017) no.2,  026022
  doi:10.1103/PhysRevD.96.026022  
  
\bibitem{Lashkari:2015hha}
  N.~Lashkari and M.~Van Raamsdonk,
  JHEP {\bf 1604} (2016) 153
  doi:10.1007/JHEP04(2016)153 
  
  \bibitem{MIyaji:2015mia}
  M.~Miyaji, T.~Numasawa, N.~Shiba, T.~Takayanagi, K.~Watanabe,
  Phys.\ Rev.\ Lett.\  {\bf 115} (2015) no.26,  261602
  doi:10.1103/PhysRevLett.115.261602.  
  
\bibitem{Kachru:2008yh}
  S.~Kachru, X.~Liu and M.~Mulligan,
  Phys.\ Rev.\ D {\bf 78} (2008) 106005
  doi:10.1103/PhysRevD.78.106005  
  

  
  \bibitem{Banerjee:2017qti}
  S.~Banerjee, J.~Erdmenger, D.~Sarkar,
  JHEP {\bf 1808} (2018) 001.  
      
  
  \bibitem{paris2009quantum}
M.~G. Paris, 
``Quantum estimation for quantum technology,''
  International Journal of Quantum Information {\bf{07}} (2009) 125.  
  
\bibitem{Bures}
Bures, D, 
``An Extension of Kakutani's Theorem on Infinite Product Measures to the Tensor Product of Semifinite w*-Algebras,''
 Transactions of the American Mathematical Society,{\bf{135}} (1969) 199-212.
 doi:10.2307/1995012  
  
  \end{thebibliography}
\end{document}